\documentclass[]{article}
\usepackage{natbib}
\usepackage{emulateapj}
\usepackage{graphicx}
\usepackage{multirow}
\usepackage{color}

\def\msun{{\rm M}_{\odot}}
\def\rsun{{\rm ~R}_{\odot}}

\def\zsun{{\rm ~Z}_{\odot}}

\begin{document}

\title{Double compact objects II: Cosmological merger rates}

 \author{Michal Dominik\altaffilmark{1}, 
         Krzysztof Belczynski\altaffilmark{1,2}, 
         Christopher Fryer\altaffilmark{3}, 
         Daniel E. Holz\altaffilmark{4}, 
         Emanuele Berti\altaffilmark{5,6}, 
         Tomasz Bulik\altaffilmark{1}, 
         Ilya Mandel\altaffilmark{7}, 
         Richard O'Shaughnessy\altaffilmark{8}, 
 }

 \affil{
     $^{1}$ Astronomical Observatory, University of Warsaw, Al.
            Ujazdowskie 4, 00-478 Warsaw, Poland  \\
     $^{2}$ Center for Gravitational Wave Astronomy, University of Texas at
            Brownsville, Brownsville, TX 78520\\
     $^{3}$ CCS-2, MSD409, Los Alamos National Laboratory, Los Alamos, NM 87545 \\
     $^{4}$ Enrico Fermi Institute, Department of Physics, and Kavli Institute
for Cosmological Physics\\University of Chicago, Chicago, IL 60637\\
     $^{5}$ Department of Physics and Astronomy, The University of Mississippi, 
	    University, MS 38677, USA \\
     $^{6}$ California Institute of Technology, Pasadena, CA 91109, USA \\
     $^{7}$ School of Physics and Astronomy, University of Birmingham, Edgbaston, B15 2TT, UK \\
     $^{8}$ Pennsylvania State University, University Park, PA \\
 }

\begin{abstract}
The development of advanced gravitational wave (GW) observatories, such as Advanced LIGO and
Advanced Virgo, provides impetus to refine theoretical predictions for what these
instruments might detect. In particular, with the range increasing by an order
of magnitude, the search for GW sources is extending beyond the ``local''
Universe and out to cosmological distances. Double compact
objects (neutron star-neutron star (NS-NS), black hole-neutron star (BH-NS) and black 
hole-black hole (BH-BH) systems) are considered to be the most promising gravitational wave 
sources. In addition, NS-NS and/or BH-NS systems are thought to be the
progenitors of gamma ray bursts (GRBs), and may also be associated with
kilonovae. In this paper we present the merger event rates of these objects as a function of cosmological 
redshift. We provide the results for four cases, each one investigating a different important evolution 
parameter of binary stars. Each case is also presented for two metallicity evolution scenarios. We 
find that (i) in most cases NS-NS systems dominate the merger rates in the local Universe, while 
BH-BH mergers dominate at high redshift; (ii) BH-NS mergers are less frequent than other sources 
per unit volume, for all time; and (iii) natal kicks may alter the observable properties of populations 
in a significant way, allowing the underlying models of binary evolution and compact object formation 
to be easily distinguished. This is the second paper in a series of three. The third paper will
focus on calculating the detection rates of mergers by gravitational wave telescopes.
\end{abstract}


\section{Introduction}
Among the potential sources of GW, the merger of double compact objects (DCOs) is
considered the most promising one for the first detection. The next generation of gravitational
wave observatories (i.e., Advanced LIGO, Advanced Virgo, KAGRA) 
will probe the Universe in search for DCO signatures at unprecedented distances,
reaching cosmological scales ($z>0.1$). 
In this paper we present predictions for DCO merger rates from isolated  (i.e.,
field population) DCO progenitors as a function of cosmological redshift. 

The distribution of binary coalescence as a function of redshift has been investigated 
by several authors. An important initial work was the investigation of the 
redshift distribution of GRBs (e.g. \cite{totani}). Preliminary work on
the importance of GW measurements of chirp mass distributions was done
by \cite{bbr}, while initial studies of the GW confusion background have
been presented in \cite{reghugh}.

In the first paper in this series (\cite{dominik}, first in the series) we investigated the sensitivity 
of DCO formation to major uncertainties of binary evolution (regarding mostly supernovae and 
common envelope episodes (CE)). We presented several models to bracket the
current uncertainty in the phenomena deciding the fate of DCO systems.
Building on this work, in the current study we present a set of four
evolutionary models. In addition to
a standard (reference) model, we have added models investigating a range of Hertzsprung gap CE donors, 
supernova (SN) explosion engines, and BH natal kicks (see Section \ref{suite} and 
Table \ref{list}). Additionally, for each model we have performed the evolutionary calculations for 11 
metallicity values, allowing us to cover the abundance of metals in Population I
and II stars (see Sections \ref{galpop} and \ref{binary}).

\begin{figure}
\includegraphics[angle=270,width=1.0\columnwidth]{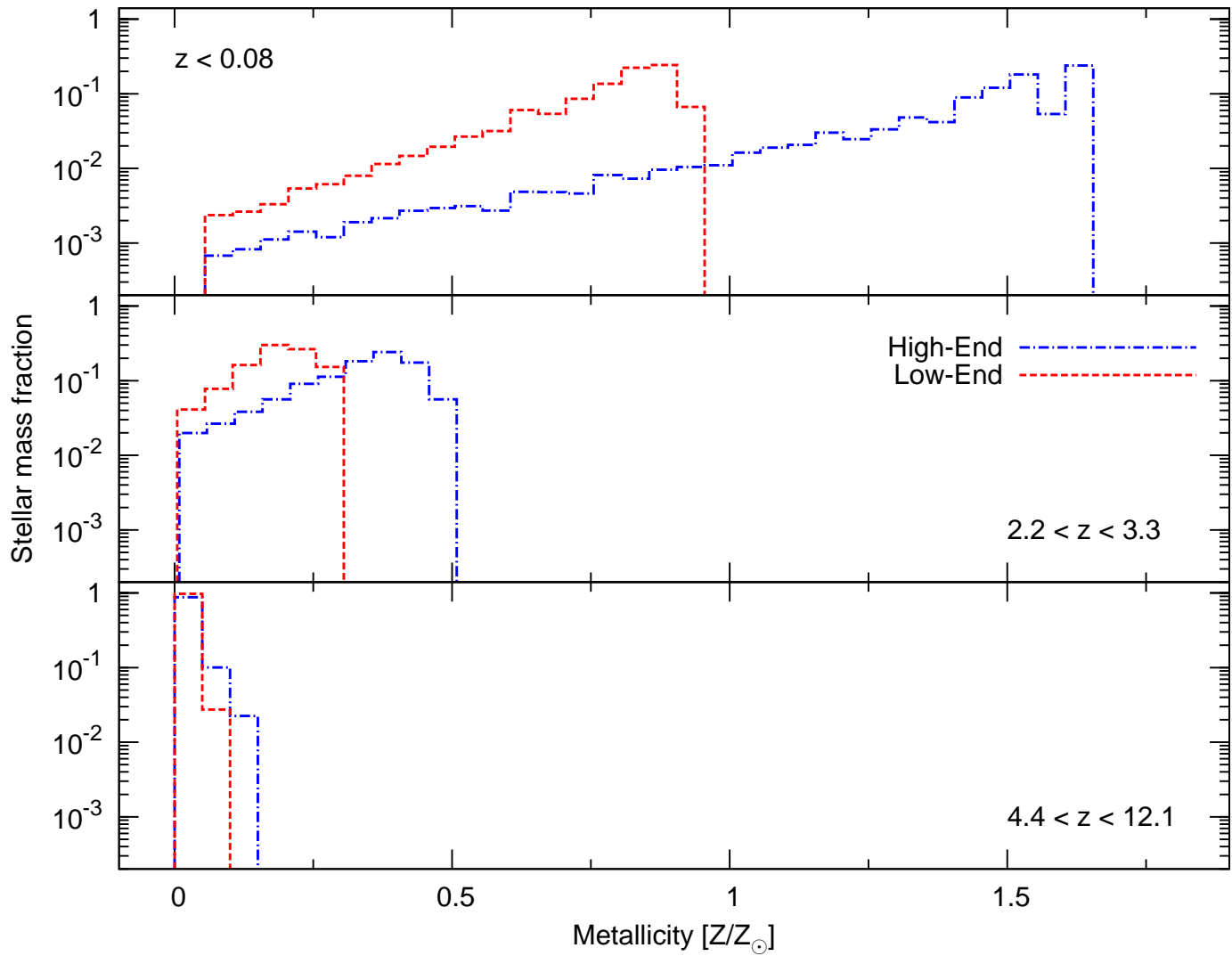}
\vspace{1.8 cm}
\caption{
Distribution of metallicity for $z<0.08$ (local Universe), $2.2<z<3.3$ (star formation peak)
and $4.4<z<12.1$ (high--redshift Universe). The y--axis shows the fraction of the total stellar 
mass in the given redshift range. The dashed and dash-dot lines represent the distributions for 
the final \textit{low--end} and \textit{high--end} metallicity profiles, respectively. The redshift 
ranges correspond to a $1$ Gyr time bin. Each distribution is normalized to unity within each redshift
range. See Section \ref{gmet} for details.
}
\label{masmet}
\end{figure}

\begin{figure}
\includegraphics[width=1.0\columnwidth]{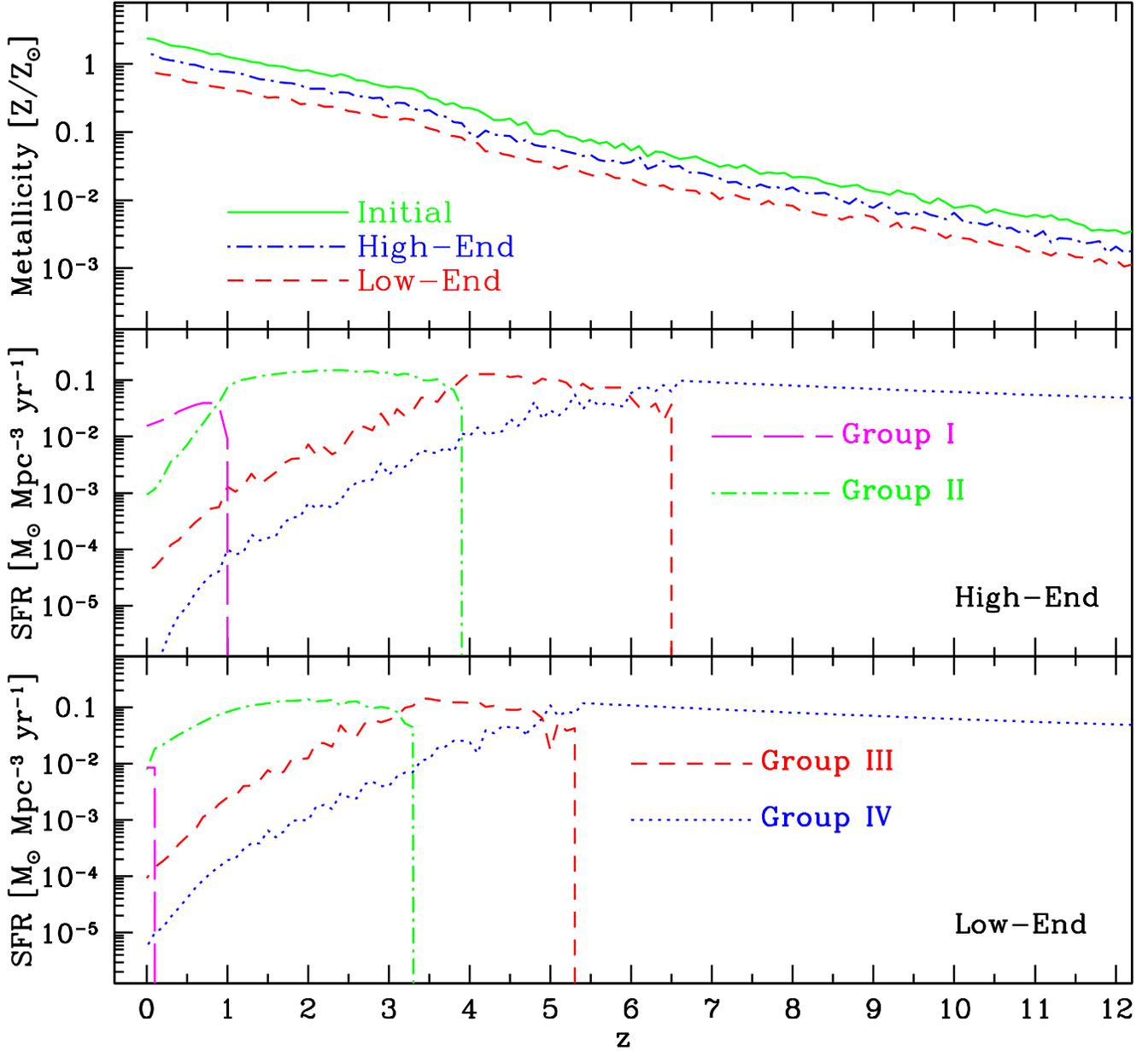}
\caption{
\textit{Top panel}. Evolution of average metallicity of galaxies with redshift. 
The dashed and dash-dot lines represents the \textit{low--end} and \textit{high--end} metallicity 
evolution scenarios, respectively. The solid line represents the \textit{initial} profile, which 
is not used in this study. See Section \ref{gmet} for details. The \textit{middle} and 
\textit{bottom} panels present the SFR divided into metallicity groups for the \textit{high--end}
and \textit{low--end} evolution scenarios. Group I contains: $1.5\zsun$ and $\zsun$; Group II:
$0.75\zsun$, $0.5\zsun$ and $0.25\zsun$; Group III: $0.1\zsun$, $0.075\zsun$ and $0.05\zsun$;
Group IV: $0.025\zsun$, $0.01\zsun$ and $0.005\zsun$. See Sections \ref{cosmo} and \ref{gmet}
for details.
}
\label{zmet}
\end{figure}

To account for the varied chemical composition of the Universe, we perform the cosmological 
calculations for two scenarios of metallicity evolution, that we will call ``low--end'' and ``high--end''
, respectively. These yield distinct rates of average metallicity growth, allowing us to ``bracket'' the 
associated uncertainties (see Section \ref{gmet}, Fig.~\ref{masmet} and Fig.~\ref{zmet}). 

In this study we investigate field stellar populations only. However, recent studies (e.g., 
\cite{kluster}) suggest that mergers in globular clusters may add a significant contribution to 
the overall coalescence rates in the Universe. In this sense, our results can be taken as 
conservative lower limits.

We present the intrinsic merger rate densities and observer frame merger rates of all three types 
of DCOs in Figures \ref{rest4high}, \ref{metform}, \ref{rest4low}, \ref{obs4high}, and \ref{obs4low}. 
Figs.~\ref{egdishigh} and~\ref{egdislow} show the BH-BH merger rate densities as
a function of the total masses of the systems. The results acquired in this study are 
available online at {\tt www.syntheticuniverse.org}.
 
\section{Stellar populations}
In this section we describe the properties of stellar populations, and their evolution
with redshift. The formalism is mostly adopted from \cite{grb}.

\subsection{Star Formation History} 
In order to determine the merger rates of DCOs we need the star formation rate (SFR). 
We adopt the formula provided by \cite{strolger}:
\begin{equation} \label{sfr}
SFR=10^9a \left(t^b e^{-t/c} +de^{d(t-t_0)/c} \right) \,\msun \textrm{yr$^{-1}$ Gpc$^{-3}$},
\end{equation} 
where $t$ is the age of the Universe (Gyr) as measured in the rest frame, $t_0$ is the
present age of the Universe ($13.47$ Gyr, see Section \ref{cosmo}) and the parameters 
have values: $a=0.182$, $b=1.26$, $c=1.865$ and $d=0.071$. The SFR described above is expressed 
in comoving units of length and time.

\subsection{Galaxy Mass Distribution}
For redshifts $z<4$ we describe the distribution of galaxy masses using a Schechter-type
probability density function, calibrated to observations~\citep{fontana}:
\begin{equation} \label{galmass}
\Phi(M_{{\rm gal,z}})=\Phi^*(z)\ln(10)a^{1+\alpha(z)}e^{-a},
\end{equation}
where $\Phi^*(z)=0.0035(1+z)^{-2.2}$, $a=M_{\rm gal} \cdot 10^{-M_{\rm z}}$ 
($M_{\rm z}=11.16+0.17z-0.07z^2$), and $\alpha(z)=-1.18-0.082z$. A galaxy mass 
is drawn from this distribution in solar units ($\msun$) and in the range 
$7 < \log(M_{\rm gal}) < 12$. Beyond redshift $z=4$ we assume no 
further evolution in galaxy mass, fixing the mass distribution 
to the value at $z=4$. This assumption reflects the lack of information on
galaxy mass distribution at high redshift. 

\subsection{Galaxy Metallicity} \label{gmet}
We assume the average oxygen to hydrogen number ratio 
($F_{\rm OH}=\log(10^{12}{\rm O/H})$) in a typical galaxy to be given by
\begin{equation} \label{meteq}
\log(F_{\rm OH})=s+1.847 \log(M_{\rm gal})- 0.08026 (\log(M_{\rm gal}))^2.
\end{equation}
As suggested 
by \cite{erb} and \cite{ynf}, the functional form of this mass-metallicity relation is 
redshift independent, with only the normalization factor $s$ varying with redshift. 
We describe the redshift dependence of galaxy metallicity using the average metallicity 
relation from \cite{pei}:
\begin{equation}
Z \propto \left\{
\begin{array}{l r}
10^{-a_2 z} & z < 3.2\\
10^{-b_1-b_2 z} & 3.2 \leq z < 5\\
10^{-c_1-c_2 z} & z \geq 3.2\\
\end{array} \right. ,
\end{equation}
which implies the evolution of $s$ with redshift:
\begin{equation}
s \propto \left\{
\begin{array}{l l}
-a_2z -1.492 & z < 3.2 \\
-b_2z -3.2(a_2-b_2)-1.492 & 3.2 \leq z < 5 \\
-c_2z -5(b_2-c_2) -3.2(a_2-b_2)-1.492 & z \geq 3.2 \\
\end{array} \right. 
\end{equation}
We assume that the oxygen abundance (used in $F_{\rm OH}$) 
correlates linearly with the average abundance of elements heavier than Helium
(encapsulated in the metallicity measure, $Z$). 

In this paper we employ two distinct scenarios for metallicity evolution with redshift in order 
to investigate the uncertainties of the chemical evolution of the Universe. The construction of 
these scenarios consists of several steps. (1.) We utilize two normalizations of 
Eq.~\ref{meteq}. In the first, provided by \cite{pei}, the coefficients are: $a_2=0.5$, $b_1=0.8$, 
$b_2=0.25$, $c_1=0.2$, $c_2=0.4$. This grants a rate of average metallicity evolution, which we label
\textit{slow}. The second, provided by \cite{ynf}, uses $a_2=0.12$, $b_1=-0.704$, $b_2=0.34$,
$c_1=0.0$, $c_2=0.1992$. It is based on ultraviolet-GALEX, SDSS, infrared-Spitzer and 
neutrino-Super Kamiokande observations \citep{hopkins}. This normalization
produces a faster rate of chemical evolution, and we label the
results \textit{fast}. At this point, for each galaxy mass value at a given redshift (Eq.~\ref{galmass}) we 
have two metallicity values (Eq.~\ref{meteq}). (2.) We then combine these (\textit{slow} 
and \textit{fast}) metallicities into a single value being an average of the
two; we label this
profile as \textit{initial}. However, this profile yields an unrealistically high number of galaxies 
with extrasolar (up to 3$\zsun$; $\zsun=2\%$ of stellar mass) metallicities at
redshift $z\sim 0$. 
(3.) In order to be consistent with observational data,  we scale down the
profile so that it agrees with the 
observed metallicities of galaxies in the local Universe (at $z\sim 0$). We explore two such ''extreme'' 
scalings resulting in a pair of final metallicity evolution
profiles. In the first, we divide the metallicity 
values from the \textit{initial} profile by a factor of $1.7$. This grants a median value of metallicity 
of $1.5\zsun$ at $z\sim 0$ (see Fig.~\ref{masmet}), which corresponds to $8.9$ in the ''12+log(O/H)'' 
formalism. This calibration was designed to match the upper $1 \sigma$ scatter of metallicities according to
\cite{yuan} (see their Fig.~2, top-right panel). We label this profile as \textit{high--end}, as it 
is the upper limit on metallicity at $z\sim 0$. In the second, we utilize SDSS observations \citep{panter}, 
from which we infer that one half of the star forming mass of the galaxies at $z\sim0$ has
$20\%$ solar metallicity, while the other 
half has $80\%$ solar metallicity. This yields a median metallicity value of $0.8\zsun$ and 
requires the division of the \textit{initial} profile by a factor of $3$. We label this profile as 
\textit{low--end}.

\subsection{Galaxy Stellar Populations} \label{galpop}
We distinguish three stellar populations:
\begin{equation}
\begin{array}{l l}
F_{\rm OH,gal} < 10^{-4} & \textrm{Population III} \\
10^{-4} \leq F_{\rm OH,gal} \leq 10^{-1} & \textrm{Population II} \\
F_{\rm OH,gal} > 10^{-1} & \textrm{Population I}
\end{array}.
\end{equation}
We choose $F_{\rm OH,gal}=10^{-4}$ as the delineation point between Population II and III stars.
A lower abundance of metals provides insufficient cooling in the collapse of gas 
clouds, and thus significantly alters the star formation for Population III stars
\citep[e.g.][]{mackey,smith}. The border point between Population II and I stars is dictated by observations 
in the Milky Way \citep[e.g.][]{binney,beers}.

We assume that the binary fraction is $50\%$: for each single star there exists one
binary. We additionally assume that all the stars within each galaxy share the same metallicity
value. The use of average metallicity seems to be appropriate since we draw a large 
($10^4$) number of galaxies (Eq.~\ref{galmass}) via Monte Carlo simulations.

\section{Binary Star Modeling} \label{binary}
We present our calculations for a set of 4 models, each differing in major input 
physics (see Table \ref{list} and the subsequent sections). For each model we use a 
grid of 11 metallicity values ($Z=0.03$, $0.02$(solar,$\zsun$), $0.015$, $0.01$, $0.005$, 
$0.002$, $0.0015$, $0.001$, $0.0005$, $0.0002$, $0.0001$) in order to accurately account for 
the average metallicity evolution of the stellar populations with redshift. 

\subsection{The {\tt StarTrack} code}
To calculate the evolution of the stellar populations we utilize the recently 
updated \citep{onthemax,massgap,dominik} {\tt Startrack} population synthesis code 
\citep{comprehensive,startrack}. This code can evolve isolated binary stars that are interacting 
in quasi-hydrostatic equilibrium from the Zero Age Main Sequence (ZAMS), through mass transfer, 
to the formation of compact objects, and to the ultimate merger of the binary components. The 
code makes use of an extensive set of formulae and prescriptions that adequately approximate more 
detailed binary evolution calculations (see \cite{hurley}).

{\tt StarTrack} allows to investigate stable and unstable mass transfers between 
the binary components. Stable transfer calculations have been calibrated on massive binaries
that are relevant to DCO formation \citep[e.g.][]{tau1999,well2001}. It is yet unknown exactly
how conservative the stable mass transfer is. \cite{dp2003} suggest that in massive binaries the 
fraction of the envelope of the donor accreted by its companion ranges between $40\%$ and $70\%$.
In our calculations we fix this value to be $50\%$ or in mathematical terms:
\begin{equation}
\dot{M_{\rm acc}}=f_{\rm a}\dot{M_{\rm don}},
\end{equation} 
where $\dot{M_{\rm acc}}$ is the accretion rate, $\dot{M_{\rm don}}$ is the mass transfer rate from
the donor and $f_{\rm a}$ is the fraction of the rate transferred (here equal to $0.5$). The remaining
mass is expelled to infinity.
The dynamically unstable mass transfers (common envelope) is calculated according 
to the energy balance formula \citep{webbink}. with the envelope binding energy parameter $\lambda$ 
adopted from \cite{chlambda}. 

Tidal interactions and their influence on eccentricity, the semi-major axis and rotation is also 
evaluated. The calculations are done with the standard equilibrium-tide, weak-friction approximation 
\citep{zahn77,zahn89}, using the formalism of \cite{hut81}. However, the code does not allow to investigate 
the influence the rotation of the components has on their internal structure.

Stellar winds are taken into account as a function of the metallicity and evolutionary stage of the 
star. This piece of physics is especially important as it has a significant impact on the masses
of remnant objects, which are the centerpiece of this study. In short, the wind mass loss rates are
divided into categories specific to the evolutionary stage of the star: O/B--type, Red Giant, Asymptotic
Red Giant, Wolf-Rayet stars and Luminous Blue Variable (LBV) stars. The magnitude of the 
rates increases with metallicity of the star except for the LBV phase. In this stage the winds are set
to be of the order of $10^{-4} \msun$yr$^{-1}$. This value was calibrated to account for the highest
mass black holes in the Milky Way $\sim 15 \msun$ (Cyg X-1 and GRS 1915). A detailed description of 
wind mass loss rates can be found in \cite{onthemax}. 

Besides stellar winds, the code also calculates changes of the angular momentum arising from gravitational 
radiation and magnetic braking. The latter is adopted from \cite{ivan2003}. 

Additionally, the utilizes the convection driven, neutrino--enhanced supernovae engines 
\citep{chrisija} to determine the properties of the remnant objects (neutron stars and black holes).

For each metallicity value in each model we evolve $2\times 10^6$ binaries, 
assuming that each component is created at the same time. Each binary system is initialized
by four parameters which are assumed to be independent. These are: primary
mass, $M_1$ (initially more massive component), mass ratio, $q=M_2/M_1$, where 
$M_2$ is the mass of the secondary component (initially less massive), the
semi-major axis, $a$, of the orbit, and the eccentricity, $e$. The mass of the primary component
is randomly chosen from the initial mass function adopted from \cite{kro1} and
\cite{kro2}:
\begin{equation} \label{imf}
\Psi (M_1) \propto \left\{
\begin{array}{l c}
M_1^{-1.3} & \quad 0.08 \ \msun \leq M_1 < 0.5 \ \msun \\
M_1^{-2.2} & \quad 0.5 \ \msun \leq M_1 < 1.0 \ \msun \\
M_1^{-\alpha} & \quad 1.0 \ \msun \leq M_1 < 150 \ \msun, \\
\end{array}
\right.
\end{equation}
where $\alpha=2.7$ is our standard choice for field populations. The choice of the upper IMF 
cutoff ($150\msun$) is justified by observations of massive stars in the Milky Way \citep{figer,oey}.
Stars are generated from within an initial mass range, with the limits  based on the targeted stellar
population. For example, studies of single neutron stars require their evolution within the range 
$8$--$20\,\msun$, while for single BHs the lower limit is $20\,\msun$. Binary evolution
may broaden these ranges due to mass transfer episodes, and we therefore set the
minimum mass of the primary to $5\,\msun$. We assume a flat mass ratio distribution, 
$\Phi(q)=1$, over the range $q=0$--1, in agreement with recent observations~\citep{kob}.
Given a value of the primary mass and the mass ratio, we obtain the mass of the secondary 
from $M_2=qM_1$. However, for the same reasons as for the primary, we don't consider binaries 
where the mass of the secondary is below $3\msun$. The distribution of initial binary 
separations is assumed to be flat in $\log(a)$~\citep{abt}, and so $\propto 1/a$, 
with $a$ ranging from values such that at ZAMS the primary fills no 
more than 50\% of its Roche lobe to $10^5 \ \rsun $. For the initial eccentricity we adopt a thermal 
equilibrium distribution \citep[e.g.][]{heggie,duq}: $\Xi(e)=2e$, with $e$ ranging from $0$ to $1$.
We find that the adopted parameters are in accordance with the most recent observations of O-star 
populations \citep{sana}.

\subsection{The model suite} \label{suite}
\subsubsection{The Standard Model} \label{smodel}
In this subsection we define a reference model for this paper.
This model is identical with the ``Standard model -- submodel B'' in the previous paper
in this series \citep{dominik}.

The list of major parameters describing the input physics of binary evolution in this 
model begins with the \textit{Nanjing} $\lambda$ \citep{chlambda} common envelope (CE) 
coefficient used in the energy balance prescription \citep{webbink}. This $\lambda$ value 
depends on the evolutionary stage of the donor, its mass at ZAMS and the mass 
of its envelope, and its radius. In addition, all of these quantities depend on metallicity,
which in our simulations changes within a broad range ($Z=10^{-4}$--$0.03$).

However, before calculating the aforementioned energy balance to determine the outcome
of the CE we check the evolutionary type of the donor star. For example, Main Sequence 
(MS) stars do not have clear core-envelope division, as the helium core is still in the 
process of being developed. Donors on the HG behave similarly,
although it remains unclear if such a division can appear on the HG or not until
later stages, like the Core Helium Burning (P. Eggleton, private
communication). In our previous work we investigated two possibilities of the CE 
outcome associated with the type of the donor star: 
an automatic (premature) merger if the donor is a HG star, regardless of the energy balance
(labeled as ``Submodel B") or allow the CE energy balance to unfold (``Submodel A''). 

The case in which we allow for potential survival of systems with HG donors
results in very high Advanced LIGO/VIRGO detection rates \citep[$\sim 8000$
  yr$^{-1}$;][]{bhkick}, exceeding even the empirically estimated rates based on
IC10 X-1 and NGC 300 X-1 \citep[$\sim 2000$ yr$^{-1}$;][]{bulik2011,cygx3}.
Therefore, we only show one model with this generous assumption on CE physics,
which leads to the most optimistic of our predictions. This model
(Optimistic CE) will be tested (and probably quickly eliminated) by the upper limits from the
Advanced LIGO/VIRGO engineering runs. For all the other models, including our reference model, 
we make the conservative assumption that none of the HG donor CE phases leads to the
formation of DCOs.

Observations suggest \citep{hobbs} that neutron stars formed in supernovae receive natal kicks, 
with velocities drawn from a Maxwellian distribution with $\sigma=265\,\mbox{km}/\mbox{s}$. We employ these
findings in our simulations, and extend them so that black hole natal kicks match this distribution
as well. However, it is possible that some matter ejected during the explosion
will not reach the escape velocity, and will thus fall back on the remnant object, potentially stalling 
the initial kick. To account for this, we modify the Maxwellian kicks by the amount of matter falling 
back on the newly formed compact object:
\begin{equation} \label{vkick}
V_{\rm k}=V_{\rm max}(1-f_{\rm fb}),
\end{equation}
where $V_{\rm k}$ is the final magnitude of the natal kick, $V_{\rm max}$ is
the velocity drawn from a Maxwellian kick distribution, and $f_{\rm fb}$
is the fallback factor describing the fraction of the ejecta returning to the object. 
The values of $f_{\rm fb}$ range between 0--1, with 0 indicating no fallback/full kick 
and 1 representing total fallback/no kick \citep[a ``silent supernova'',
  e.g.][]{mirabel}. We label this the ``constant velocity'' formalism. An alternative approach
is the ``constant momentum'' one, where the kick velocity is inversely proportional to the mass
of the remnant object. In general, constant velocity provides larger natal kicks on average than
constant momentum resulting in more frequent disruptions of binaries, especially for systems with BHs. 
Therefore, we choose the ``constant velocity'' formalism over the ``constant momentum'' as it provides 
a more conservative limit on the number and therefore merger rates of systems containing BHs.

This model also utilizes the ``Rapid'' convection driven, neutrino enhanced supernova 
engine \citep{chrisija}. It allows for a successful explosion without the need for the 
artificial injection of energy into the exploding star. In this scenario the explosion 
starts from the Rayleigh-Taylor instability and occurs within the first $0.1$--$0.2\,\mbox{s}$.
For low mass stars ($M_{\rm zams} \lesssim 25 \msun$) the result is a very strong (high 
velocity kick) supernova, which generates a NS. For higher mass stars a BH is formed 
through a direct collapse (failed supernova). This engine, incorporated into binary evolution,
successfully reproduces the mass gap \citep{massgap} observed in Galactic X-ray binaries
\citep{mg1,mg2}.

The list of major physical parameters used in this and subsequent models is given in 
Table \ref{list}. More details on the physics described above can be found in \cite{dominik}.

\subsubsection{Variations on the standard model} \label{variations}

The uncertainties in the CE and the SN engine argue for exploring a range of
input physics beyond that in the standard model described in the previous subsection.
In this subsection we present three additional models which we have found to encapsulate the
full range of possible binary evolutions. All subsequent models use the same input physics as 
the reference model, except for the parameters described below.

\textit{Optimistic Common Envelope}. In this model we allow HG stars to be CE donors
(see Section \ref{smodel}). When the donor initiates the CE phase the energy
balance determines the outcome. This model is identical to the ``standard 
model -- submodel A'' from our previous paper in this series \citep{dominik}.

\textit{Delayed SN}. This model utilizes the ``Delayed'' supernova engine instead of the 
Rapid one. The Delayed is also a convection driven, neutrino enhanced engine, but is 
sourced from the standing accretion shock instability (SASI), and can produce an
explosion as late as $1\,\mbox{s}$ after bounce. The Delayed engine produces a continuous 
mass spectrum of compact objects, from NSs, through light BHs, to massive BHs 
(see \cite{massgap}). This model is identical to the ``Variation 10 -- submodel B''
model from our previous paper in this series \citep{dominik}.

\textit{High BH kicks}. In this model the BHs receive full natal kicks.
The newly formed BH acquires a velocity drawn from a Maxwellian distribution 
(see Section \ref{smodel}) regardless of the fallback factor $f_{\rm fb}$ (see Eq.~\ref{vkick}).
This model is identical to the ``Variation 8 -- submodel B'' model in our previous paper
in this series \citep{dominik}. 

\section{Cosmology calculations} \label{cosmo}
We utilize a flat cosmology with $H_0=70\,\mbox{km}\,\mbox{s}^{-1}\,\mbox{Mpc}^{-1}$, $\Omega_M=0.3$,
$\Omega_{\Lambda}=0.7$, and $\Omega_k=0.0$. The relationship between redshift and time
is given by:
\begin{equation}
t(z)=t_H\int^{\infty}_{z} \frac{dz'}{(1+z')E(z')},
\end{equation}
where $t_H=1/H_0=14$ Gyr is the Hubble time \citep[e.g.][]{hogg} and 
$E(z)=\sqrt{\Omega_M(1+z)^3+\Omega_k(1+z)^2+\Omega_{\Lambda}}$.

The comoving volume element $dV$ is given by:
\begin{equation} \label{dvdz}
dV(z)=\frac{c}{H_0}\frac{D_c^2}{E(z)}d\Omega dz,
\end{equation}
where $c$ is the speed of light in vacuum, $d\Omega$ is the solid angle, and $D_c$ is
the comoving distance given by :
\begin{equation}
D_c(z)=\frac{c}{H_0} \int^z_0 \frac{dz'}{E(z')}.
\end{equation}

There are a series of steps to calculate the rates of events, as we now
describe.
We employ time as our reference coordinate and start by creating
time bins across the entire history of Universe, each bin $100$ 
Myrs wide, from $0.13$ Gyrs (birth) to $13.47$ Gyrs (today). At the center of each bin
we evaluate the star formation rate according to Eq.~\ref{sfr}. For the redshift value
corresponding to the center of a given time bin we generate a Monte Carlo sample of $10^4$ 
galaxies (a number sufficient to produce a smooth distribution) with masses drawn from the 
distribution given in Eq.~\ref{galmass}. For each time bin we obtain a total mass 
of galaxies $M_{\rm gal,tot}$.
For each galaxy we then estimate its average metallicity using 
Eq.~\ref{meteq}. We assume that {\em all} stars within a given galaxy have identical
metallicity as obtained from Eq.~\ref{meteq}. Since we draw a large number of galaxies
in each time bin, and each galaxy has its own mass, and therefore is
described by its own average metallicity, we end up with a distribution of metallicity 
in each time bin. This also yields
a total mass of galaxies with a specific metallicity 
($M_{\rm gal,i}$) within each time bin. We then define the fraction of the total galaxy mass 
capable of forming stellar population with a specific metallicity by
\begin{equation}
F_i=\frac{M_{\rm gal,i}}{M_{\rm gal,tot}}.
\end{equation}
However, because we use a finite number of metallicity points in our simulations (see Section \ref{binary})
we need to extrapolate our results in order to account for the continuous spectrum given by Eq.~\ref{meteq}. 
Therefore, the metallicity points are extended into bins delineated by the average value of 
neighbouring points. For example, given the set of points $Z=0.01,0.015,0.02$, the value $Z=0.015$ 
now corresponds to a bin that extends from $0.0125$ to $0.0175$. The border points $Z=0.0001$ and $Z=0.03$ 
extend to lower and higher values, respectively, to cover the rest of the spectrum. 

Population synthesis provides us with a representative sample of DCOs.
The formation of a single DCO within a time 
bin corresponds to a fraction, $f_{\rm fr}$, of the total formation rate:
\begin{equation}
f_{\rm fr}(t)=\frac{F_i}{M_{\rm sim}}SFR(t),
\end{equation}
where $M_{\rm sim}$ is the total mass in our population synthesis calculations (see Section 
\ref{binary}). Repeating this calculation of $f_{\rm fr}$ for each metallicity yields a total 
formation rate, $f_{\rm fr,tot}$, within a given time bin.

We now need to know the delay time until merger, $t_{\rm del}$, for each
DCO formed. The delay time is defined
as the interval between the formation of the progenitors of a DCO and the coalescence of two
compact objects. For each DCO originating from a specific metallicity we randomly choose a birth 
point, $t_0$ (ZAMS), within each time bin. We then propagate the system forward in time
toward its merger using the delay time:
\begin{equation}
t_{\rm mer}=t_0+t_{\rm del}.
\end{equation}
As long as we consider DCOs with $t_{\rm mer}<t_H$, and as long as the width of the time bins throughout the 
time line is constant, the formation rate ($f_{\rm fr}$) of a DCO from a given bin translates into 
a merger rate in a later bin, propagated forward in time by $t_{\rm del}$. Repeating
the above calculations for every time bin yields a total density of
rest frame merger events, $n_{\rm rest}(t)$, in units of
Gpc$^{-3}\,\mbox{yr}^{-1}$. In other words
\begin{equation}
n_{\rm rest}(t)=\sum^N_i f_{{\rm fr},i}(t-t_{\rm del}),
\end{equation}
where $i$ sums over each representative DCOs.

\section{Results} \label{results}
We now provide results from our four models, presenting the intrinsic merger
rate densities and the observer frame merger rates, given by
\begin{equation}
n_{\rm obs}(<z)=4\pi \int^{z}_0 \frac{n_{\rm rest}}{1+z'}\frac{dV}{dz'}dz' \quad [\rm{yr}^{-1}],
\end{equation}
with $dV/{dz}$ given by Eq.~\ref{dvdz}, integrated over the solid angle $d\Omega$
(hence the factor of $4\pi$).
In the case of the standard/reference model (details in Section \ref{smodel}) we explain the
general redshift behavior of all three types of DCOs (NS-NS, BH-NS, and BH-BH) and compare the
reference model for two scenarios of metallicity evolution (\textit{high--end}
and \textit{low--end}). For our three variations (Optimistic CE, Delayed SN, and High BH
kicks, Section \ref{variations}) we investigate deviations from the
reference model, again incorporating our different metallicity evolution scenarios.

\subsection{Standard Model} \label{smodelres}
{\bf NS-NS}. As shown in Fig.~\ref{rest4high} the intrinsic merger rate densities of double
neutron star systems peaks at redshift
$z\approx 1$ ($\sim 200$ yr$^{-1}$Gpc$^{-3}$). As a general rule, the merger rates
of all types of DCO are directly related to the star formation rate. However, for
a given SFR value the formation efficiency of different DCO may vary. In other words,
the proportions of NS-NS, BH-NS, and BH-BH systems may differ beyond
the regime set by the IMF \citep[e.g.][]{dominik}. For example, NS-NS systems are on
average efficiently created in high metallicity environments (see Fig.~\ref{metform}).
When combined with the peak of the SFR at $z\sim 2$ (average \textit{high--end} metallicity is
$\sim 0.4\zsun$, see Fig.~\ref{zmet}), high metallicity NS-NS formation efficiency is enhanced,
thus creating the profile shown on Fig.~\ref{rest4high}. What is characteristic for this profile is the
''hump'' that arises at $z\sim 1.6$, approaching from high redshifts.
As can be seen in Fig.~\ref{metform}, this shape is dominated by mergers originating from $0.75\zsun$
environments. The reason for this increase in merger rate densities, when transiting from $0.5\zsun$
environments (higher redshifts) to higher metallicity ones (lower redshifts), is a consequence of the applied
CE approach. Within the framework of the \textit{Nanjing} CE treatment adapted for the {\tt Startrack}
code, the binding energy of the CE decreases at the $0.5$--$0.75\zsun$ boundary, allowing for the survival
of a larger number of NS-NS progenitors.

By comparison, for the \textit{low--end} metallicity profile
the NS-NS systems dominate the rates only up to $z\approx 0.5$
(Fig.~\ref{rest4low}). This is a consequence of the adopted metallicity evolution scenario.
Specifically galaxies of a given metallicity are shifted to
lower redshift when compared with the \textit{high--end}
scenario, causing a shift of the NS-NS systems also to lower redshifts.

As shown in Fig.~\ref{obs4high}, in the observer frame the systems dominate the merger rates up to
redshift $z\approx 2.4$. However, decreased metallicity for the \textit{low--end} case
shifts this point to $z\approx 0.6$ (Fig.~\ref{obs4low}).

{\bf BH-NS}. For the \textit{high--end} metallicity evolution, the rest frame
merger rate densities for BH-NS systems shown in Fig.~\ref{rest4high} peaks at a
value of $\sim 50\,\mbox{Gpc}^{-3}\,\mbox{yr}^{-1}$ at redshift $z\sim 3$. However,
the merger rate efficiency drops for low ($z\sim 0$) and high ($z\sim 6$)
redshifts. This is because of properties of the progenitor masses.
For metallicities $\sim \zsun$ the bulk of the progenitors masses
are in the range $45$--$60\,\msun$ for the primary component, and
$22$--$32\,\msun$ for the secondary. Pairs of progenitors outside these ranges
are unlikely. The upper mass limit delineates between BH-NS and BH-BH systems; crossing it
results in the formation of the latter systems instead of the former. The lower mass
limit is set by a similar phenomenon, only this time through BH-NS/NS-NS formation. Progenitors of these
systems for metallicities a factor of $\sim 10$ lower than $\zsun$ must have lower masses on average, primarily
because of the decreased stellar wind mass losses. Otherwise the binary would retain enough mass to
form a BH-BH system or go through a terminal CE event. Therefore, the mass ranges for BH-NS
progenitors for $Z\sim 10\%\zsun$ are: $20$--$50\,\msun$ for the primary and $12$--$25\,\msun$ for the
secondary. Given that in this mass range the Initial Mass Function (IMF) scales as $M^{-2.7}$ (where
$M$ is the mass of the progenitor) there are more BH-NS progenitors available at moderately low
metallicity than at higher values. This, in turn, translates into increased merger rates arising from these
environments. Decreasing the metallicity to $\sim 1\% \zsun$ decreases the masses of the
progenitors even further due to the same wind effects. However, in this case the BH progenitors are
closely approaching their lower mass limit ($\sim 20\msun$), which leaves a narrow mass range:
$20$--$25\msun$ for the primary and $18$--$22\msun$ for the
secondary. There are fewer progenitors in these mass
ranges when compared to the previous case, and therefore we find a lower
merger rate. Overall, the BH-NS merger rates peak originates from systems
being created at moderate metallicities (see Fig.~\ref{metform}).

{\bf BH-BH}. For these systems the intrinsic merger rate has a peak-plateau at a
rate of $\sim 300$--$400\,\mbox{Gpc}^{-3}\,\mbox{yr}^{-1}$ at $z\sim 4$--$8$,
for the \textit{high--end} case.
The low metallicity galaxies abundant at high redshifts are efficient black hole factories 
(see Fig.~\ref{egdishigh} and \ref{egdislow}). This also means that adopting the \textit{low--end} 
metallicity scenario will allow for more BH-BH systems to form at lower redshifts, when compared to the 
\textit{high--end}. Additionally, environments with low amounts of metals favor massive BHs. For 
example, the most massive BH-BH system acquired in this model consisted of a $62\,\msun$ and a $74\,\msun$ 
BH pair. These systems originate from the extremely low metallicity environments ($Z=0.0001$). We find that 
such systems merge up up until redshift $z\sim 3$ and $z\sim 2$ for the \textit{high--end} and \textit{low--end}
metallicity evolution models, respectively.
However due to statistical uncertainties these redshift values may be even lower. These massive systems
originate through the standard BH-BH formation channel. As an instructive example, we detail
the formation scenario of a $8.3$--$5.8\,\msun$ BH-BH, for $Z=0.005$ -- a typical system for the average 
metallicity acquired in our study: {\bf t=0 Myr.} The components start with masses $32\,\msun$ and 
$25\,\msun$ for the primary and secondary, respectively and an orbital separation $a=995\,\rsun$. 
{\bf t=6.7 Myr.} The primary, after becoming a HG star, expands and initiates a mass transfer through 
Roche lobe overflow (RLOF). The transfer continues until the primary loses almost all of its hydrogen 
envelope and becomes a Wolf-Rayet star with $10\,\msun$ (the secondary component has $35\,\msun$). 
The orbital separation prior to RLOF was $a=1000\,\rsun$ and $a=1600\,\rsun$ after. 
{\bf t=7.0 Myr.} The primary explodes as a supernova, forming a $7.8\,\msun$ BH. The orbital separation
after the explosion was $a=1760\,\rsun$ {\bf t=8.7 Myr.} The secondary ($34\,\msun$) initiates a CE phase 
and becomes a Wolf-Rayet star with $11\,\msun$ as a result of the outcome (the primary gained 
$\sim 0.5\,\msun$) The orbital separation prior to the CE was $a=1780\,\rsun$ and $a=2.6\,\rsun$ after. 
{\bf t=9.4 Myr.} The secondary undergoes a SN explosion and becomes a $5.8\,\msun$ BH. The orbital 
separation prior to the explosion was $a=2.8\,\rsun$ and $a=3\,\rsun$ after.
{\bf t=26 Myr.} The coalescence of a $8.3\,\msun$--$5.8\,\msun$ BH-BH system occurs. This example is 
illustrated by a diagram in Fig.~\ref{diagram}. 

On a side note, the formation of the most massive BH-BH systems on close orbits may be questionable.
The progenitors of the aforementioned $62\,\msun$-$74\,\msun$ BH-BH system are massive stars
($140\,\msun$--$150\,\msun$ at ZAMS). A recent theoretical study by \cite{yusof2013} suggests that such objects
($150\,\msun$--$500\,\msun$) will not increase in size significantly during their evolution. Therefore,
it is more likely for such binaries to bypass the CE phase and avoid the reduction of orbital 
separation. This in turn will prevent the resulting BH-BH system from merging within Hubble time.

In the observer frame, BH-BH systems begin to dominate the merger rates at $z\sim 2$. For the
\textit{low--end} case this happens closer to $z\sim 1$.

\subsection{Optimistic CE} \label{relaxed}
In this model we relax one of the conditions on CE survivability. Specifically, Hertzsprung
gap donors are now allowed to undergo full energy balance calculations. In the standard model, CEs
with HG donors resulted in an immediate merger, terminating further binary evolution. This has
been shown to have a significant impact on the number of DCOs, altering the merger rates
by orders of magnitude \citep{rarity,nasza}. When HG donors survive CE,
their numbers naturally increase, as do their merger rates.

{\bf NS-NS}. When compared with the standard model, \textit{high--end} case, the intrinsic merger
rate of NS-NS systems peaks at higher redshift ($z\sim 3$) and at higher values
($\sim 1000$ Gpc$^{-3}$ yr$^{-1}$).
The shift in the peak towards higher redshifts is associated with the systems having shorter
delay times on average, which allows them to merge more quickly after formation. As expected, the decrease
of average delay times for NS-NS systems is caused by the new CE condition. In the standard model
the only surviving binaries were those that did not initiate the CE while the potential donor was an
HG star. In order to prevent a rapidly expanding HG star from overfilling its Roche lobe these
binaries had to have a significant initial separation, which resulted in relatively large delay
times. In this model the CE phase with an HG donor is allowed, so initial separation is no longer
such a crucial issue. Therefore, binaries with smaller initial
separations are able (if they have sufficient orbital energy) to survive and form NS-NS systems.
This results in shorter delay times. In the
\textit{low--end} case the same mechanism causes the peak to shift towards $z\sim 2$.

In the observer frame the merger rate of NS-NS systems is a few times higher
than in the standard model for both \textit{high--end} and \textit{low--end} metallicity evolutions.
In the former case NS-NS systems dominate the merger rates up to $z\lesssim0.5$.
In the latter case they are always sub-dominant compared to BH-BH systems.

{\bf BH-NS}. The binaries forming these systems usually undergo two CE events in their lifetime,
due to their relatively high initial mass ratios \citep[for details see][]{dominik}. The two CEs
reduce the initial separation, which makes the relaxed CE condition much less relevant than for the
NS-NS case mentioned above. The result is that there are no significant changes
in the intrinsic merger rate density for BH-NS systems. As in the standard model, the mergers of BH-NS
systems are the rarest of all types of DCOs. This is true for both of the metallicity scenarios.

{\bf BH-BH}. These systems do not experience two CE events, unlike the BH-NS
systems, and therefore they do not reduce their initial separations as efficiently.
The peak of the intrinsic merger rate density shifts slightly towards
lower redshift ($z\sim 4$, \textit{high--end}) when contrasted with the
reference model (see Fig.~\ref{rest4high}). This is because of the
effect of metallicity on the outcome of the CE phase.
The larger the fraction of metals in a star, the bigger its radius
\citep[e.g.][]{hurley}. This effect is particularly strong during the HG phase. Therefore, high
metallicity BH progenitors are more likely to initiate CE on the HG.
In the standard model this is not allowed and such systems are removed from the population. However, here
we relax this condition, and as a consequence we add more BH-BH systems originating from higher
metallicities (see Figs.~\ref{egdishigh} and~\ref{egdislow}). For the \textit{low--end} case
this results in a peak-plateau between redshifts $3<z<4$. This is because of the higher
metallicities appearing at lower redshifts when compared with the \textit{high--end} case.

In the observer frame the BH-BH systems start to dominate the merger rates at $z\approx 0.5$ in
the \textit{high--end} case. For the \textit{low--end} case these DCOs are
always primary mergers.

\subsection{Delayed SN}
In this model we change the supernova explosion engine with respect to the standard model.
The standard model uses the Rapid engine, which yields a gap between $2$--$5 \msun$ in the 
masses of the resulting compact objects. Here we utilize the Delayed scenario (for details 
see Section \ref{variations}). The main feature of this engine is that it produces a continuous 
mass spectrum of remnant objects \citep{massgap}. As suggested by \cite{kreidberg}, the presence
of the mass gap feature may be a result of systematic errors arising from misinterpretation of
the BH binary light curve analyses. The resulting errors in estimating the inclination of the 
binary may shift low mass BHs from the gap. The distinction
between the two engines is clearly visible on Fig.~\ref{egdishigh} and Fig.~\ref{egdislow}. 
The minimal total mass for this model is $\sim 5\msun$. Such a system is composed of two BHs of
$2.5\msun$ each ($2.5\msun$ being the delineation between upper NS and lower BH mass). For other 
models the minimal BH mass is $\sim 5\msun$, thus yielding a minimal total mass $\sim 10\msun$. 
However, the supernova engine effects do not play a significant role on the merger rates of any 
type of DCOs.

\subsection{High BH kicks}
Here, we employ full natal kicks (as measured for NSs) just on BHs (see Section
\ref{variations}). This is performed regardless of the amount of fallback (see Eq.~\ref{vkick}).
The kicks for NS-NS systems remain unchanged, as does their population with respect to the standard model.

In this variation the velocity of the natal kick acquired upon BH formation will disrupt many
binaries that would otherwise (in the standard model) form coalescing BH-NS or
BH-BH systems, as is clearly visible in Fig.~\ref{rest4high}. In consequence the NS-NS systems will
dominate the merger rates in the observer frame.

In addition, the full natal kick will affect the most massive BHs. In the standard model,
stars with masses $M_{\rm zams} > 40 \msun$ would collapse directly into a black hole after
the SN explosion; with no asymmetric ejecta, they do not receive a kick ($f_{\rm fb}=1$, Eq.~\ref{vkick}). 
However, in this model these stars receive a maximum velocity kick, and thus often disrupt the system. 
As a consequence the probability of the formation and eventual merger of the most massive BH-BH systems 
is lowered significantly, which can be seen on the bottom panel of Figs.~\ref{egdishigh} and~\ref{egdislow}.

\section{Summary \& Discussion}
We have performed a series of cosmological calculations for four populations of DCOs.
Each population was generated with different input physics for describing binary
evolution and compact object formation. The first model (standard) utilizes the
current state-of-the-art description of physical mechanisms governing
DCOs. In particular, it uses a \textit{Rapid} explosion engine, which
yields results accurately describing the mass distribution of X-ray binaries (see Section
\ref{smodel} and references therein). Another major improvement in the model is the realistic
treatment of the common envelope parameter $\lambda$, which now depends on the evolutionary
stage, radius, mass, metallicity, etc. of the donor star. The three subsequent models explore
alternative outcomes of binary evolution, and the resulting properties of remnants.
The mechanisms investigated in these models are: the sensitivity of the CE outcome to the
type of donor, the Delayed SN explosion mechanism, and the natal kick survivability
of DCOs containing BHs (see Section \ref{variations}). Additionally, for each model we
have created a grid of 11 metallicities to account for the chemical evolution throughout
the lifetime of the Universe. We present both the intrinsic and the
observer frame merger rates as a function of redshift.

The variation in the rates of our different binary systems as a function of redshift
depends upon metallicity, as well as common envelope and supernova physics.
In this paper we have studied how these impact the rates for different types of
DCOs. Here we review our main findings.

\begin{figure}
\includegraphics[width=1.0\columnwidth]{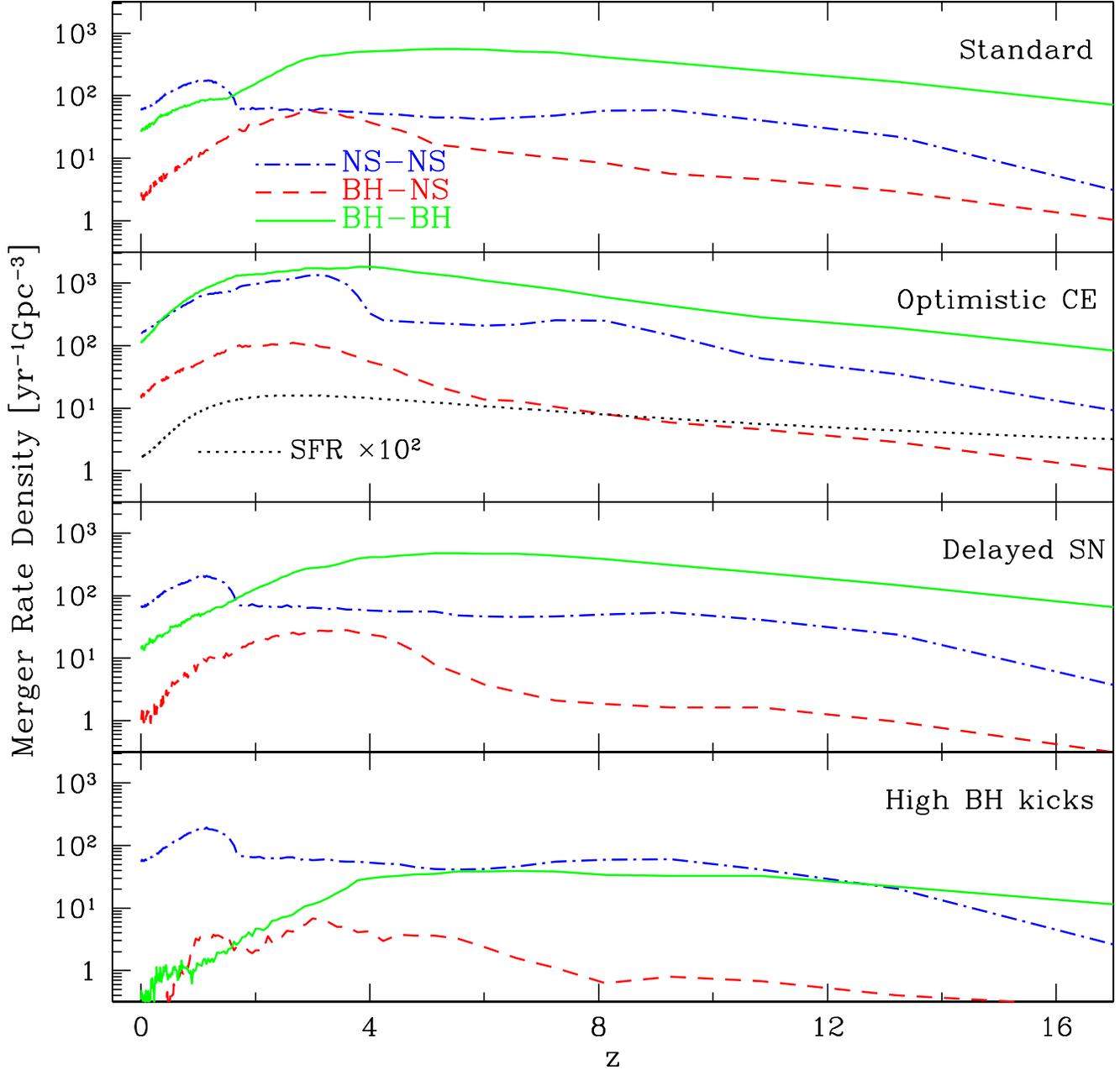}
\caption{DCO merger rate densities in the rest frame (intrinsic), for \textit{high--end} metallicity.
Each panel shows a different model, as listed (for details see Section
\ref{suite}). The dash-dot, dashed, and solid lines represent NS-NS,
BH-NS, and BH-BH systems, respectively. The dotted line in the second panel from the top represents
the star formation rate (see Eq.~\ref{sfr}) multiplied by a factor of $100$
for clarity; it is in units of $\msun/100\,\mbox{Mpc}^{-3}\,\mbox{yr}^{-1}$.
This figure demonstrates: (i) a clear domination of NS-NS systems for the
standard model for $z\lesssim 1.6$, as these systems merge copiously in the relatively metal-rich,
local Universe; (ii) significantly increased merger rates for the Optimistic CE model, where
CE events on the Hertzsprung gap are allowed; and (iii) a drastic drop in rates for the High BH
kick model.
}
\label{rest4high}
\end{figure}

\begin{figure}
\includegraphics[angle=270,width=1.0\columnwidth]{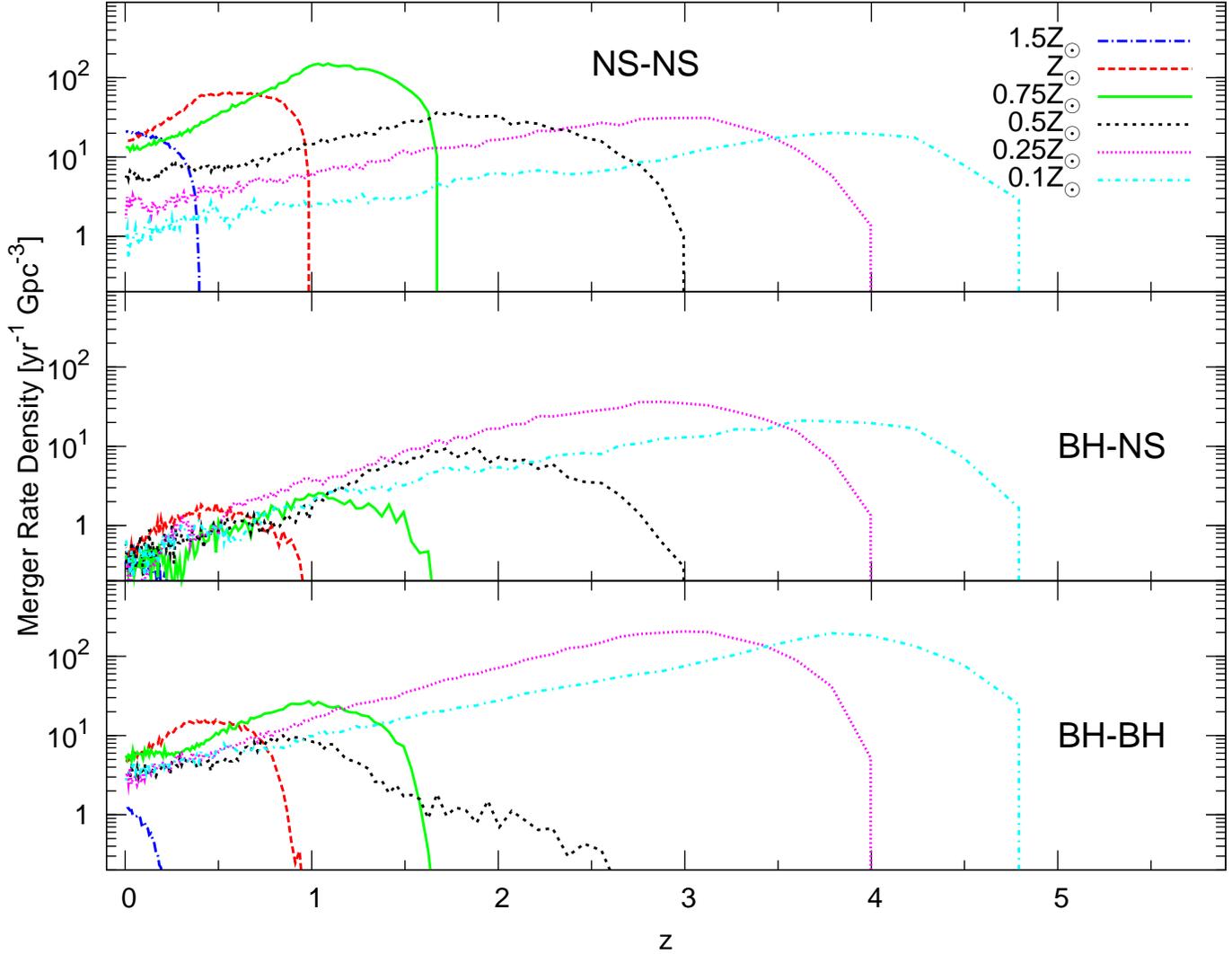}
\vspace{1.8 cm}
\caption{
Merger rates density versus redshift and metallicity: this example figure shows how the total
merger rate density as a function of redshift is built up from systems formed in different star
forming conditions. Each line indicates the contribution to the total merger rate density from 
each of the listed metallicity bins, for the standard model. The sum of these curves (and the
low metallicity curves not shown in this plot) reproduce the total merger rate density shown in
the top panel in Figure \ref{rest4high}.
\textit{Top}, \textit{middle} and \textit{bottom} panels present 
rate densities for NS-NS, BH-NS and BH-BH systems, respectively. The peak of the NS-NS systems merger 
rate density in the restframe is composed mostly of system created in $0.75\zsun$ environments. For 
BH-NS systems the peak arises from $0.25\zsun$ environments and for BH-BH systems for $0.25\zsun$--
$0.1\zsun$. 
}
\label{metform}
\end{figure}

\begin{figure}
\includegraphics[width=1.0\columnwidth]{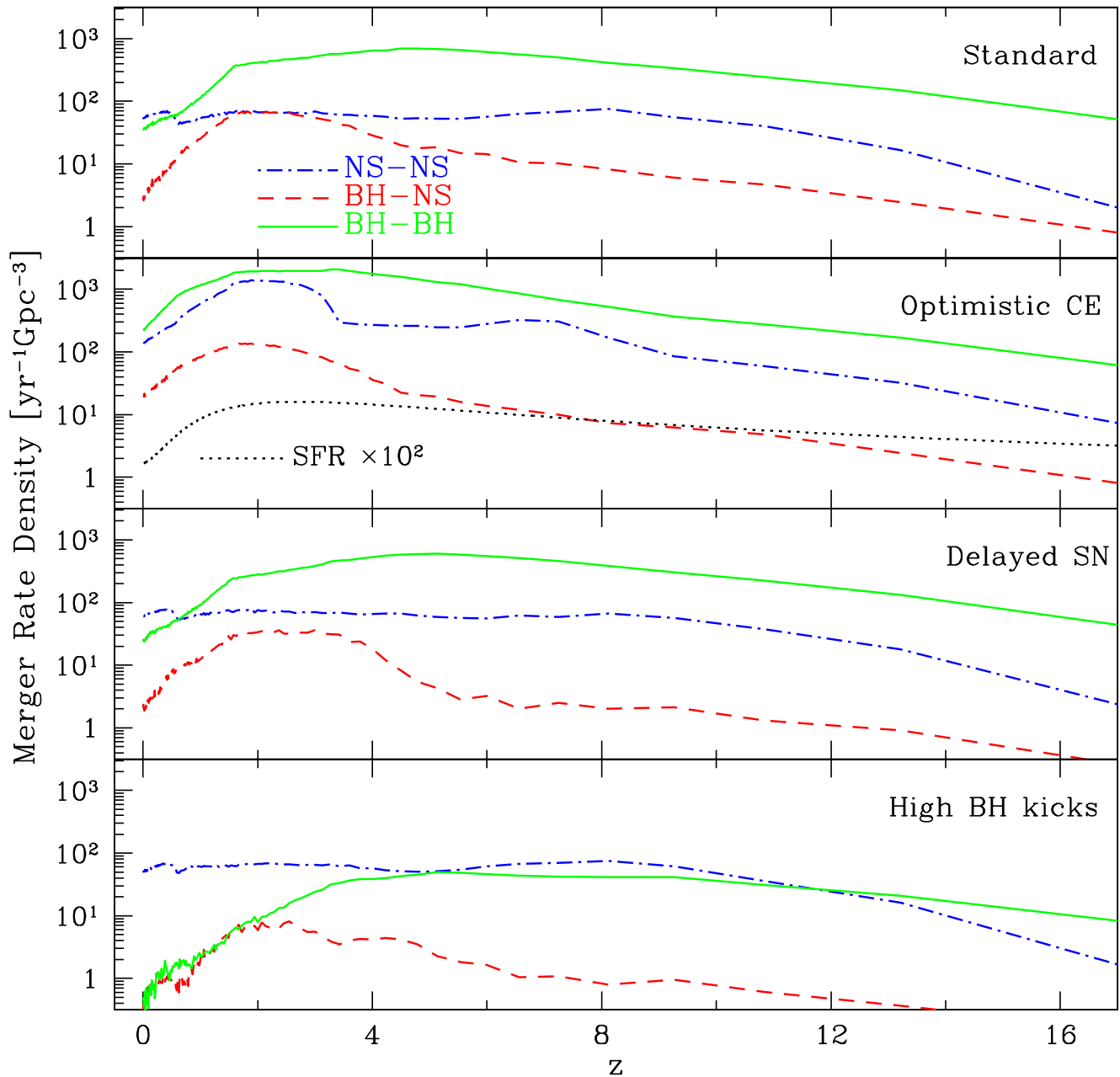}
\caption{The DCO merger rate density in the restframe (intrinsic), \textit{low--end} metallicity. 
For low redshifts ($z<2$) the low metallicity decreases the merger efficiency of NS-NS systems. 
}
\label{rest4low}
\end{figure}

\begin{figure}
\includegraphics[width=1.0\columnwidth]{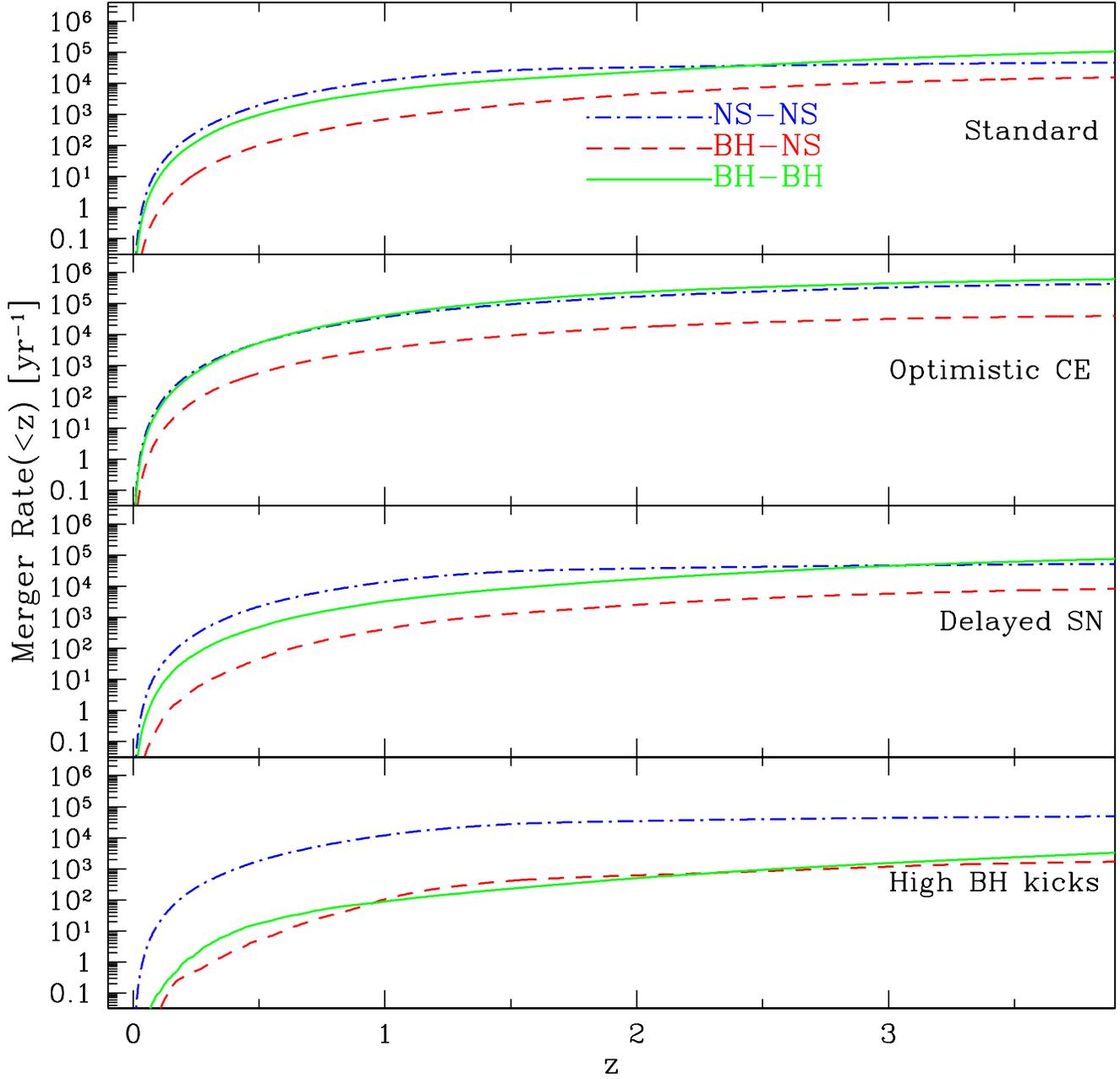}
\caption{The cumulative merger rates in the observer frame, \textit{high--end} metallicity. 
Panels are organized as in Fig.~\ref{rest4high}.
For the standard model the merger rates of NS-NS systems dominate until redshift
$z\approx 2.4$. For the Optimistic CE model the merger rates of NS-NS and BH-BH systems are very similar
until redshift $z\sim 1$, where BH-BH systems take over. For the Delayed SN model this happens at redshift
$z\sim 3$. For the High BH kicks model the NS-NS systems dominate
the merger rates throughout all redshifts.
}
\label{obs4high}
\end{figure}

\begin{figure}
\includegraphics[width=1.0\columnwidth]{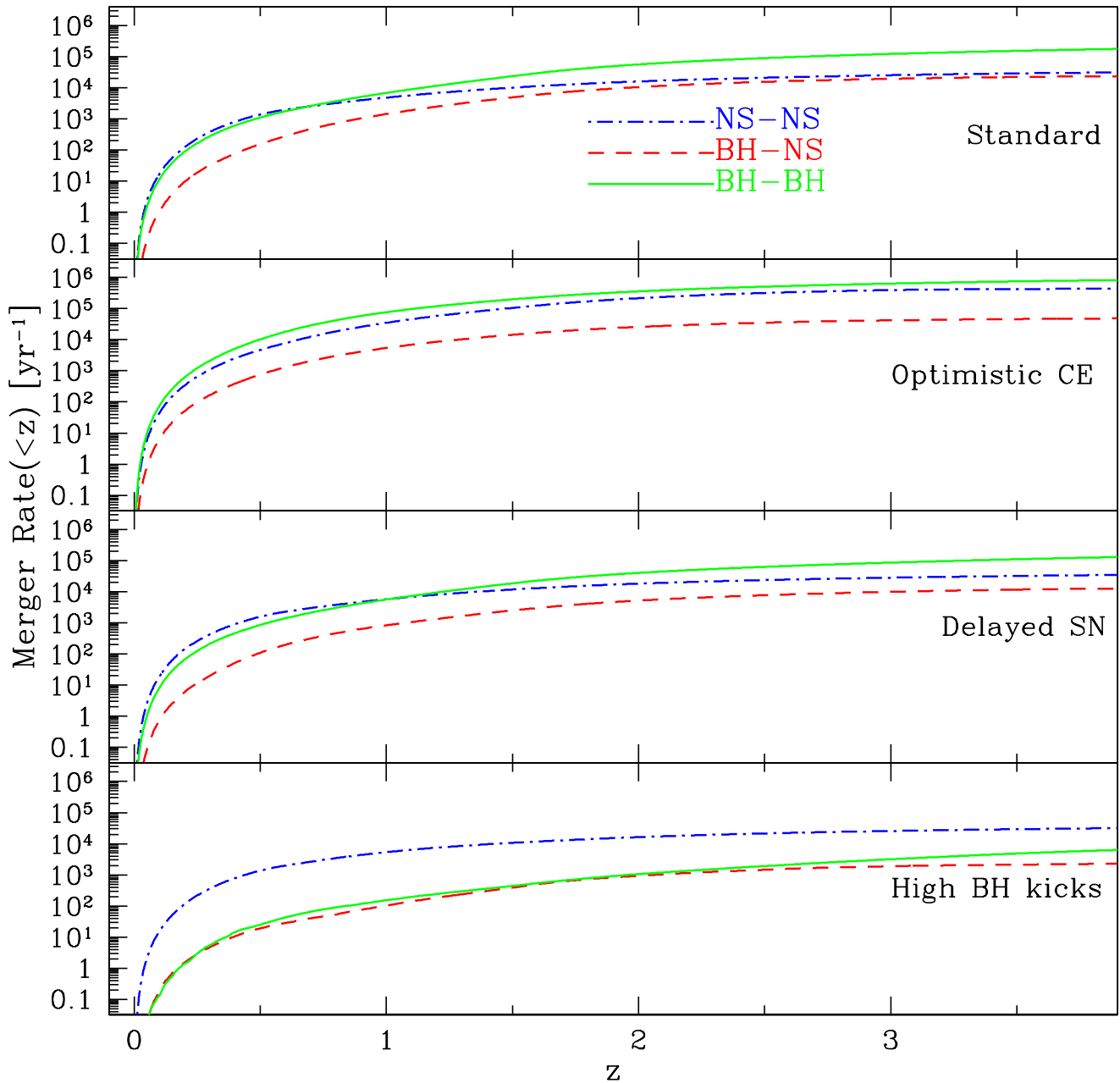}
\caption{The cumulative merger rates in the observer frame, \textit{low--end} metallicity. 
Low metallicity causes an overall decrease in merger rates of NS-NS systems when compared to 
the \textit{high--end} case (see Fig.~\ref{obs4high}).
}
\label{obs4low}
\end{figure}

We find that NS-NS systems merge most efficiently at low redshifts ($z\lesssim 1$;
see Figs.~\ref{rest4high} and~\ref{rest4low}), where metallicities become relatively high
($\sim 0.5\zsun$). However, in the case of the Optimistic CE model the merger rate densities
peak at higher redshifts ($z\sim 2$--$3$). This results from relaxing the condition
for the termination of binaries initiating a CE with a Hertzsprung gap donor.
This optimistic CE treatment enriches the merging population with systems with short
merger times. As a result the overall number of NS-NS systems increases and, due to the shorter
merger times, these systems coalesce earlier (see Sections \ref{variations} and \ref{relaxed}).

BH-NS systems merge most infrequently in all but one of the models. The exception
is the Full BH Kicks model, where full natal kicks are applied to BH
remnants. The kicks eliminate binaries containing BHs
from the populations by disrupting them. However, this doesn't affect BH-NS systems, as strongly
as BH-BH systems since they contain only one BH. In general, the low merger rates of BH-NS
systems arise from their unique mixed nature. Forming two different compact objects
in a single binary generally requires the masses of the progenitors to be significantly separated.
This plays an important role at first contact between the components, since if
the mass ratio of the binary is larger than $2$--$3$ the otherwise stable mass transfer
through Roche lobe overflow may become a CE event. These episodes often cause
a premature merger and eliminate further binary evolution. Another important
factor in making the BH-NS systems small in numbers is that the progenitors don't
have a large range of masses to draw from. The upper
limit is set by the binary containing enough mass to form a BH-BH system
instead, while the lower limit is set by not having enough mass and instead
forming a NS-NS system.

\begin{figure}
\includegraphics[width=1.0\columnwidth]{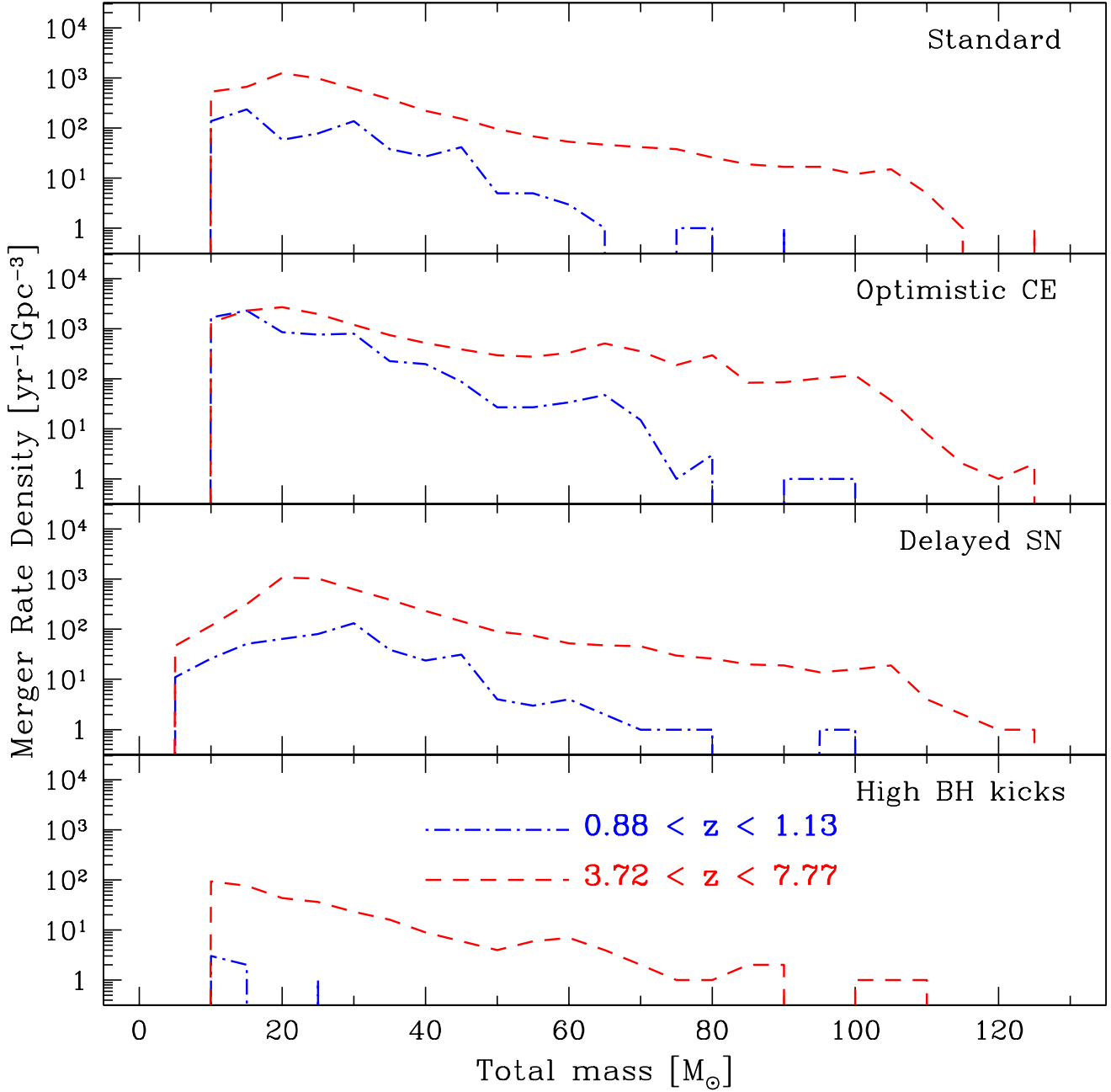}
\caption{Distribution of total mass for BH-BH system for all models, \textit{high--end} metallicity. 
The distribution is presented for BH-BH merging in two redshift ranges; one clustered around 
$z=1$ and one around $z=5$ (each range spans 1 Gyr, the corresponding limiting redshifts 
are given on the bottom panel). The redshift values are chosen arbitrarily for illustrative purposes.
Note that as the redshift decreases so does the maximum total mass of the system.
}
\label{egdishigh}
\end{figure}

\begin{figure}
\includegraphics[width=1.0\columnwidth]{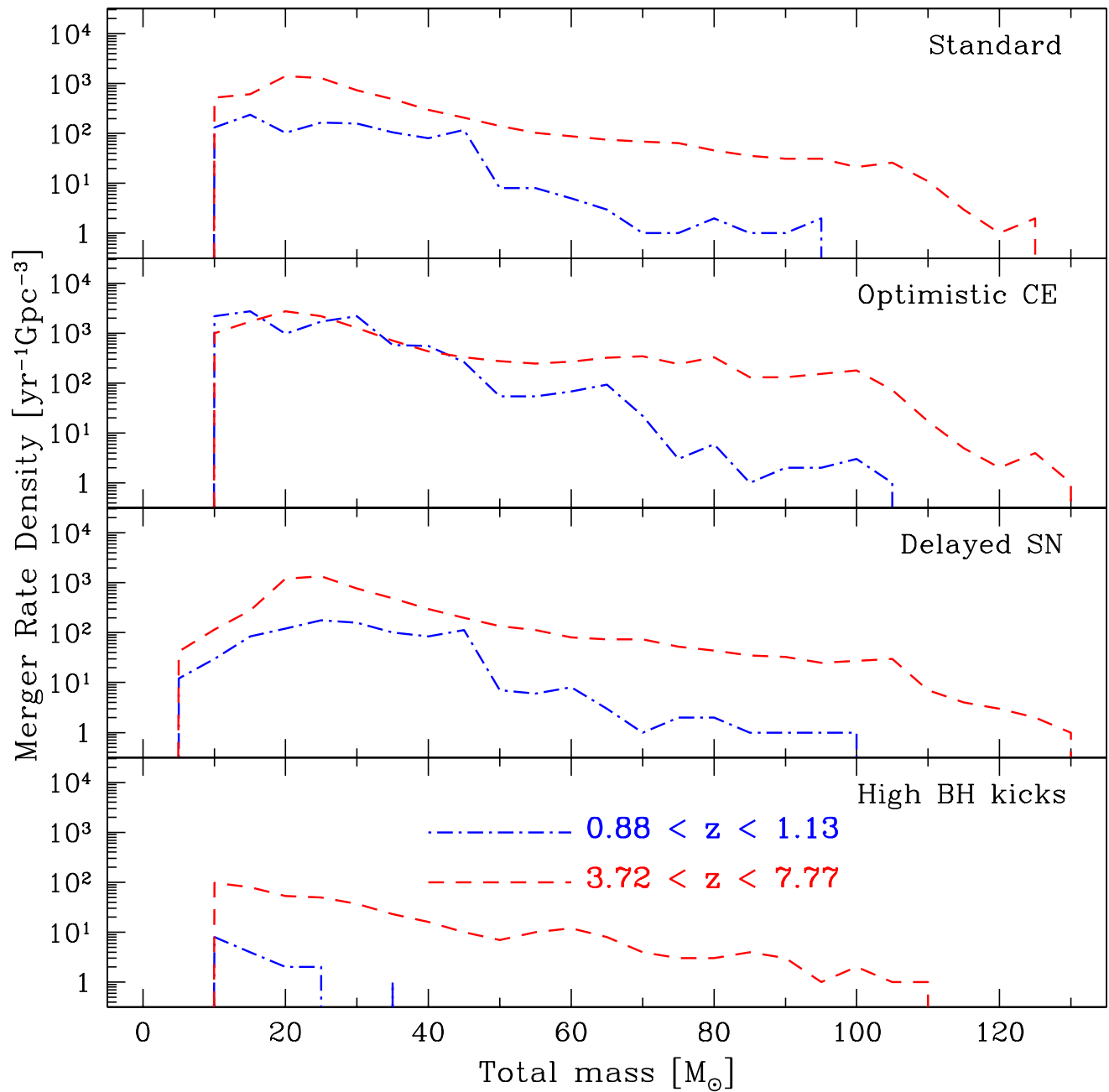}
\caption{Distribution of total mass for BH-BH system for all models, \textit{low--end} metallicity. 
}
\label{egdislow}
\end{figure}

\begin{figure}
\includegraphics[width=0.9\columnwidth]{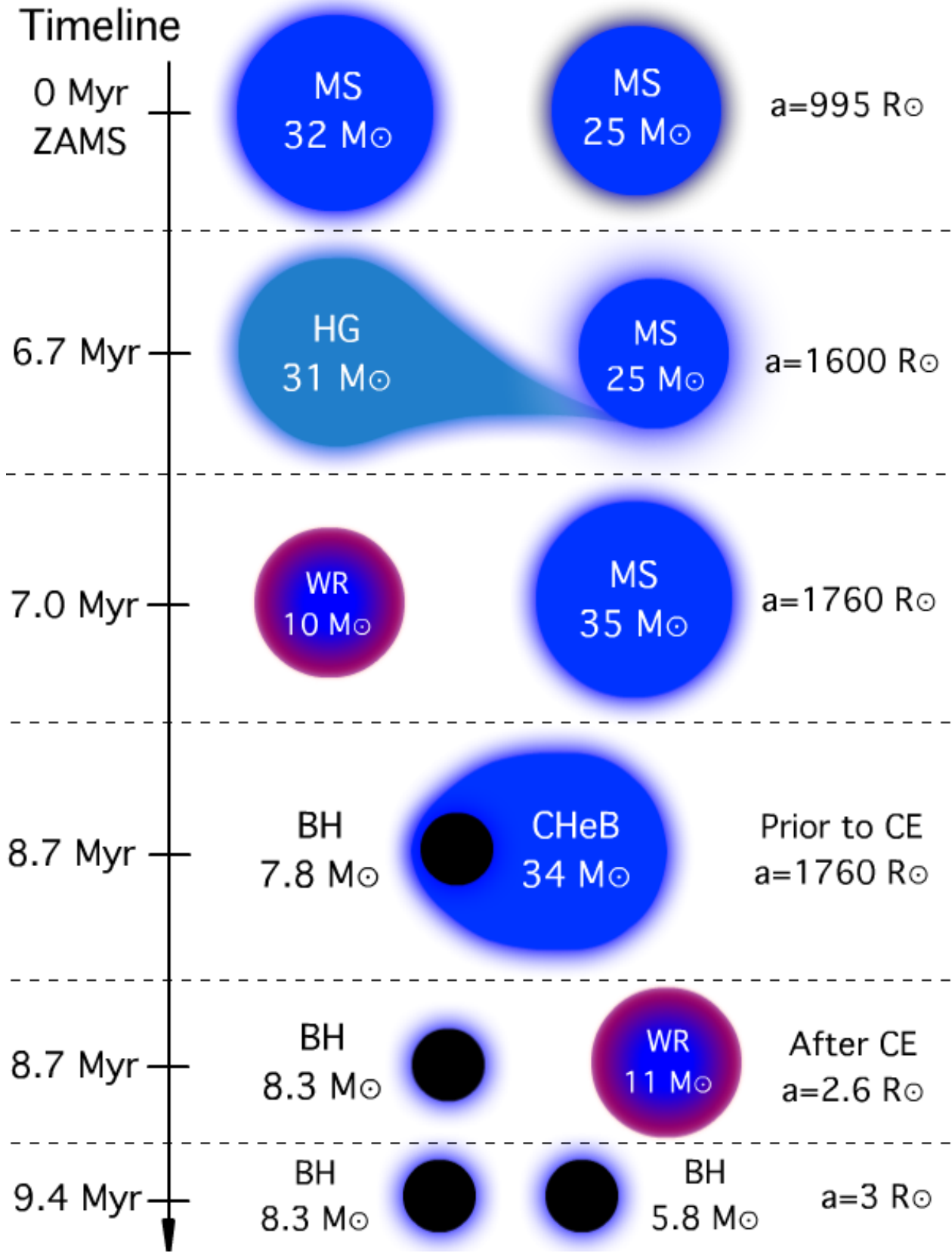}
\caption{An evolutionary diagram illustrating the example in Section \ref{smodelres}, BH-BH paragraph.
MS--Main Sequence, HG--Hertzsprung Gap, CHeB--Core Helium Burning, WR--Wolf-Rayet. From the top:
\textit{I panel}: Progenitors at Zero Age Main Sequence; \textit{II panel}: Non-conservative, stable mass
transfer from a Hertzsprung gap donor (primary) to the companion; \textit{III panel}: A WR star prior 
to a SN explosion (primary) and a rejuvenated companion; \textit{IV panel}: CE event with a CHeB donor
(secondary) and a BH accretor (primary); \textit{V panel}: The binary immediately after the CE;
\textit{VI panel}: The formation of a BH-BH system.
}
\label{diagram}
\end{figure}

For BH-BH systems, the highest merging efficiency occurs earlier in the
Universe when compared with other DCOs ($z \sim 4$--$6$). This arises from the fact that these
systems form most efficiently at the lowest metallicities. For any of the two
scenarios of metallicity evolution, the Optimistic CE model blurs this trend. In this case the population
is enriched by BHs, which originated from high metallicity environments (see Section
\ref{relaxed}). Another interesting case is the model with High BH Kicks, where BH-BH systems are
efficiently disrupted by natal kicks throughout the lifetime of the Universe. This is
clearly visible on the bottom panel of Figs.~\ref{rest4high} and~\ref{rest4low}. The
kicks affect high mass systems the most. As a consequence of the full natal kicks, the formation
and merger rates for BH-BH systems in low metallicity galaxies (higher redshifts) are reduced
significantly, and this effect is even more dramatic for high metallicity environments (lower redshifts;
see Figs.~\ref{egdishigh} and~\ref{egdislow}). The High BH kicks model produces a difference between
the merger rates in the observer frame of BH-BH and NS-NS systems that is
roughly 100 times larger than within the standard model. This may be a promising avenue for
determining the magnitude of the natal kicks imparted to BHs during their formation.

Since (only) NS-NS systems have been observed, we can use observed rates to put constraints on our
models. The NS-NS merger rates in each of our models, at $z\sim 0$, fit within the
observational limits for NS-NS systems in the Milky Way:
$34.8$--$2204\,\mbox{yr}^{-1}\,\mbox{Gpc}^{-3}$~\citep{kimkal}, using the galaxy density
$\rho_{\rm gal}=0.0116\,\mbox{Mpc}^{-3}$.
\cite{petrillo} used the observed rate of short GRBs to calculate the merger rates of
NS-NS and BH-NS systems, since these systems are thought to be the progenitors
of short GRBs. The resulting merger rates of DCOs (NS-NS + BH-NS) in
the local Universe ranges between $500$ and
$1500\,\mbox{Gpc}^{-3}\,\mbox{yr}^{-1}$. At $z\sim 0$ our models find a NS-NS
merger rate of $\sim 100\,\mbox{Gpc}^{-3}\,\mbox{yr}^{-1}$, with a BH-NS rate
lower by a factor of $\sim10$. However, the authors of the
aforementioned study state that their results are sensitive primarily to the poorly constrained beaming
angle of the colimated emission from short GRBs. They used a beaming
angle of $\sim 20$ deg, while to match our rate the beaming
angle would have to be $\sim 50$ deg (see Fig.~3 therein).
In our previous study \citep{dominik}, we found one model that would reproduce the merger rates of
NS-NS + BH-NS from \cite{petrillo} ($\sim 900\,\mbox{Gpc}^{-3}\,\mbox{yr}^{-1}$ at
$\zsun$). It is the model described by fully conservative mass transfer episodes and optimistic
CE description (labeled ''Variation 12 -- submodel A'').

Additional constraints may be provided by observing the potential electromagnetic
signatures, other than GRBs,
of DCO mergers. One example is the optical/radio afterglow of the GRB, which can
be detected even if the GRB itself is not seen (an ``orphan afterglow''). Another
possibility is a ``kilonova'', resulting from the ejection of matter from a
neutron star. Since this matter may be enriched in heavy elements through
the r--process, the resulting radioactive decay may generate observable light,
thereby providing a promising electromagnetic counterpart to the gravitational wave emission
\citep{metzger,piran,barnes}.

Finally, it will be interesting to investigate how statistical ensembles of GW observations could constrain
properties of compact binary populations and of their formation scenarios
\citep[see e.g.][]{mandel09,oshaughnessy12,gerosa}.

\acknowledgements

We thank Alexander Heger for a helpful discussion on pair--instability supernovae.
We would also like to thank the N. Copernicus Astronomical Centre in Warsaw, Poland, and
the University Of Texas, Brownsville, TX, for providing computational
resources. DEH acknowledges support from National Science Foundation CAREER grant PHY-1151836. 
KB and MD acknowledge support from MSHE grant N203 404939 and N203 511238, NASA Grant NNX09AV06A to 
the UTB, Polish Science Foundation Master 2013 Subsidy and National Science Center DEC-2011/01/N/ST9/00383.
TB was supported by the DPN/N176/VIRGO/2009 grant. EB acknowledges support from National Science Foundation 
CAREER Grant No. PHY-1055103. Work by CLF was done under the auspices of the National Nuclear Security
Administration of the U.S. Department of Energy under contract No. DE-AO52-06NA25396. 

\clearpage

\bibliographystyle{aa}
\bibliography{b1}

\clearpage

\begin{deluxetable}{c l}
\tablewidth{330pt}
\tablecaption{Summary of Models\tablenotemark{a}}
\tablehead{Model & Description}
\startdata
Standard        &  $\lambda=$\textit{Nanjing}/physical, BH kicks: decreased, SN: Rapid \\
&                  HG CE donors not allowed \\  
&\\
Optimistic CE      &    HG CE donors allowed \\
& \\
Delayed SN      &  Delayed supernova engine  \\
& \\
High BH kicks   &  Full kicks of BHs\\
\enddata
\label{list}
\tablenotetext{a}{
All parameters for a given model, except the ones given, remain as in the Standard
model. See Section \ref{suite} for details.
}
\end{deluxetable}
\clearpage

\end{document}